\documentclass[12pt]{iopart}
\input{amssym}
\usepackage{epsfig}
\global\arraycolsep=2pt

\newcommand{\ben}{\begin{equation*}}
\newcommand{\een}{\end{equation*}}
\newcommand{\bean}{\begin{eqnarray*}}
\newcommand{\eean}{\end{eqnarray*}}
\newcommand{\bnabla}{\mbox{\boldmath{$\nabla$}}}
\newcommand{\nn}{\nonumber}
\newcommand{\be}{\begin{equation}}
\newcommand{\ee}{\end{equation}}
\newcommand{\bea}{\begin{eqnarray}}
\newcommand{\eea}{\end{eqnarray}}
\newcommand{\bm}[1]{\mbox{\boldmath{$#1$}}}

\begin{document}

\title{$\mathcal{PT}$-Symmetric Quantum Electrodynamics---$\mathcal{PT}$QED}
\author{Kimball A Milton$^1$, In\'es Cavero-Pel\'aez$^2$, 
Prachi Parashar$^1$, K~V~Shajesh$^1$, and Jef Wagner$^1$}

\address{
$^1$ Oklahoma Center for High Energy Physics, and
H. L. Dodge Department of Physics and Astronomy
University of Oklahoma, Norman, OK 73019-2061 USA\\
$^2$ Laboratoire Kastler Brossel, UPMC, ENS, CNRS,
University Paris 6, Case 74,
F-72525 Paris Cedex 05, France}

\begin{abstract}
The construction of $\mathcal{PT}$-symmetric quantum electrodynamics is
reviewed.  In particular, the massless version of the theory in $1+1$
dimensions (the Schwinger model) is solved.  Difficulties with unitarity
of the $S$-matrix are discussed.
\end{abstract}

\section{$\mathcal{PT}$QED}
Quantum electrodynamics (QED) is by far the most successful physical
theory ever devised \cite{Gabrielse:2006gg}. However, although its
reach includes all of atomic physics, there are a myriad of phenomena
that we do not understand.  Thus it is essential to explore alternative
theories, in the hope that we may be able to describe aspects of the world
that are as yet not under our understanding.

One very promising new approach to quantum theories are those included
under the rubric of non-Hermitian theories, in particular theories
in which invariance under the combined operation of space and time reflection
$\mathcal{PT}$ replaces mathematical Dirac Hermiticity in order to guarantee
unitarity of the theory. (For a recent review, see \cite{Bender:2007nj}.)
Little work, however, has been done on applying this idea to quantum field
theory.  This paper represents our continuing effort to develop a 
$\mathcal{PT}$-symmetric version of quantum electrodynamics, in the hope
that a consistent theory, with a unitary $S$-matrix, results that may
eventually find physical applications in nature.

\subsection{Transformation properties}
At the first International Workshop on Pseudo-Hermitian Hamiltonians
in Quantum Physics \cite{Milton:2003ax}
a $\mathcal{PT}$-symmetric version
of quantum electrodynamics was proposed.
A non-Hermitian but $\mathcal{PT}$-symmetric electrodynamics
is based on the assumption of novel transformation properties of the
electromagnetic fields under parity $\mathcal{P}$ transformations, that is,
\be
\mathcal{P}:\quad {\bf E\to E},\quad {\bf B\to- B}, \quad {\bf A\to A},
\quad A^0\to-A^0,
\label{1}
\ee
just the statement that
 the four-vector potential is assumed to transform as an axial
vector.  Under time reversal $\mathcal{T}$, 
the transformations are assumed to be conventional,
\be
\mathcal{T}:\quad {\bf E\to E},\quad {\bf B\to- B}, \quad {\bf A\to -A},
\quad A^0\to A^0.
\label{2}
\ee
Fermion fields are also assumed to transform conventionally.
We use the metric $g^{\mu\nu}=\mbox{diag}(-1,1,1,1)$.

\subsection{Lagrangian and Hamiltonian}
The Lagrangian of the theory then possesses an imaginary coupling
constant in order that it be invariant under the product of these
two symmetries:
\be\mathcal{L}=
-\frac14F^{\mu\nu}F_{\mu\nu}-\frac12\psi\gamma^0
\gamma^\mu\frac1\rmi\partial_\mu\psi-\frac{m}2\psi\gamma^0\psi
+\frac{\rmi}2\psi\gamma^0\gamma^\mu eq\psi
A_\mu.
\label{3}
\ee
Here, because we are discarding Hermiticity as a physical requirement,
it is most appropriate to use a ``real'' field formulation, where
correspondingly the (antisymmetric, imaginary) charge matrix $q=\sigma_2$
appears. Furthermore, $\gamma^0\gamma^\mu$ is symmetric and $\gamma^0$
is antisymmetric.
 In the radiation (Coulomb) gauge $\bnabla\cdot\mathbf{A}=0$, 
the dynamical variables are $\mathbf{A}$ and $\psi$, and the canonical
momenta are $\pi_{\mathbf{A}}=-\mathbf{E}^T$, $\pi_\psi=\frac\rmi2\psi$,
where $T$ denotes the transverse part,
and so the relation between the Hamiltonian and Lagrangian densities
are
\be
\mathcal{H}=E^2+\mathbf{E}\cdot\bnabla A^0+\frac\rmi2\psi\dot\psi-
\mathcal{L}.\ee
 Then, if integrate by parts and use
$\bnabla \cdot \mathbf{E}=\rmi j^0$, we find that
the corresponding Hamiltonian is
\be
H=\int(\rmd \mathbf{r})
\left\{\frac12(E^2+B^2)+\frac12\psi\left[\gamma^0\gamma^k\left(
\frac1\rmi\nabla_k-{\rmi}
eqA_k\right)+m\gamma^0\right]\psi\right\}.
\label{4}
\ee
We can also obtain this same Hamiltonian from the stress tensor
\be
t^{\mu\nu}=-\frac\rmi4\psi\gamma^0\left(\gamma^\mu\partial^\nu+\gamma^\nu
\partial^\mu\right)\psi+
F^{\mu\lambda}F^\nu{}_\lambda-\frac\rmi2\left(j^\mu A^\nu
+j^\nu A^\mu\right)+g^{\mu\nu}\mathcal{L}.
\ee

\subsection{Current density}
The electric current appearing in both the Lagrangian and Hamiltonian
densities, 
\be j^\mu=\frac12\psi\gamma^0\gamma^\mu eq\psi,
\ee
 transforms
conventionally under both $\mathcal{P}$ and $\mathcal{T}$:
\numparts
\bea
\mathcal{P}j^\mu({\bf x},t)\mathcal{P}=\left(\begin{array}{c}
j^0\\-{\bf j}\end{array}\right)(-{\bf x},t),\label{5a}\\
\mathcal{T}j^\mu({\bf x},t)\mathcal{T}=\left(\begin{array}{c}
j^0\\-{\bf j}\end{array}\right)({\bf x},-t).
\label{5b}
\eea
\endnumparts  This just reflects the normal transformation properties
of the fermion fields.
\subsection{Equal-time commutation relations}
We are working in the Coulomb gauge, $\bm{\nabla}\cdot{\bf A}=0$, so the
nonzero canonical equal-time commutation relations are
(the fermion index $a$ includes both the Dirac and charge indices)
\numparts
\bea
\{\psi_a({\bf x},t),\psi_b({\bf y},t)\}=\delta_{ab}\delta({\bf x-y}),
\label{6a}\\
{}[A_i^T(\mathbf{x}),E_j^T(\mathbf{y})]=-\rmi\left[\delta_{ij}-\frac{\nabla_i
\nabla_j}{\nabla^2}\right]\delta(\mathbf{x-y}).
\label{6b}
\eea
\endnumparts
\be
\bm{\nabla}\cdot \mathbf{A}^T=\bm{\nabla}\cdot \mathbf{E}^T=0.
\ee

\subsection{The $\mathcal{C}$ operator}
As for quantum mechanical systems, and for scalar quantum field
theory, we seek a $\mathcal{C}$ operator in the form
\be
\mathcal{C}=\rme^{\mathcal{Q}} \mathcal{P},
\label{7}
\ee
where $\mathcal{P}$ is the parity operator.  $\mathcal{C}$ must
satisfy the properties
\numparts
\bea
\mathcal{C}^2=1,\label{8a}\\
{}[\mathcal{C},\mathcal{PT}]=0,\label{8b}\\
{}[\mathcal{C},H]=0.\label{8c}
\eea
\endnumparts

From the first two equations we infer
\be
\mathcal{Q}=-\mathcal{P}\mathcal{Q}\mathcal{P},\label{9a}\ee
and because $\mathcal{PT}=\mathcal{TP}$,
\be
\mathcal{Q}=-\mathcal{T}\mathcal{Q}\mathcal{T}.\label{9b}
\ee
The third equation \eref{8c} allows us to determine $\mathcal{Q}$ 
perturbatively.
If we separate the interaction part of the Hamiltonian from the free part,
\be
H=H_0+eH_1,\label{10}
\ee
and assume a perturbative expansion of $\mathcal{Q}$:
\be \mathcal{Q}=e\mathcal{Q}_1+e^2 \mathcal{Q}_2+\dots,\label{11}
\ee
the first contribution to the $\mathcal{Q}$ operator is determined by
\be
[\mathcal{Q}_1,H_0]=2H_1.\label{12}
\ee
The second correction commutes with the Hamiltonian,
\be
[\mathcal{Q}_2,H_0]=0.\label{13}
\ee
Thus we may take
\be
\mathcal{Q}=e\mathcal{Q}_1+e^3 \mathcal{Q}_3+\dots,\label{14}
\ee
which illustrates a virtue of the exponential form.
The $\Or(e)$ term was explicitly computed for four-dimensional
QED in 2005 \cite{Bender:2005zz}.

\section{($\mathcal{PT}$QED)$_2$}
However, the above perturbative construction of $\mathcal{C}$ fails for
2-dimensional $\mathcal{PT}$-symmetric QED.  We are here discussing
the Schwinger model 
\cite{Schwinger:1962tn,Schwinger:1962tp,brown,Lowenstein:1971fc}.
In two dimensions, the only
nonzero component of the field strength tensor is $F^{01}=E$, and the
Hamiltonian of the system is 
\be
H=\int\rmd x\left[\frac12 E^2-\rmi 
j_1A_1-\frac{\rmi}2\psi \gamma^0\gamma^1\partial_1\psi
+\frac{m}2\psi\gamma^0\psi\right].\ee

As before, we choose the radiation gauge because it is most physical:
\be
\bnabla\cdot \mathbf{A}=\partial_1A_1=0,\ee
and then the Maxwell equation becomes
\be
\partial_1 E_1=-\partial_1^2 A^0=\rmi j^0,\label{max}\ee
which implies the following construction for the scalar potential
\be
A^0(x)=-\frac{\rmi}2\int_{-\infty}^\infty \rmd y|x-y|j^0(y).\ee

\subsection{Construction of $E$}
Without loss of generality, we can disregard $A_1$, and then the electric
field is
\be
E(x)=\frac{\rmi}2\int_{-\infty}^\infty \rmd y\,\epsilon(x-y)j^0(y),\ee
with
\be
\epsilon(x-y)=\left\{\begin{array}{cc}
1,&x>y,\\0,&x=y,\\-1,&x<y.\end{array}\right.\ee
Thus the electric field part of the Hamiltonian is
\bea
\int \rmd x\, \frac12 E^2
&=&-\frac18\int \rmd x\,\rmd y\,\rmd z\, 
\epsilon(x-y)\epsilon(x-z)j^0(y)j^0(z)\nn\\
&=&-\frac18L Q^2+\frac14\int \rmd y\,\rmd z j^0(y)|y-z|j^0(z),\eea
where $L$ is the infinite ``length of space'' and the total charge is
\be
Q=\int \rmd y\, j^0(y).\ee  As this is a constant, we may disregard it.

\subsection{Form of Hamiltonian}
Thus we obtain the form found (for the conventional theory)
years ago by Lowell Brown \cite{brown}:
\be
H=\frac14\int \rmd y \,\rmd z j^0(y)|y-z|j^0(z)
-\int \rmd x\left\{
\frac{\rmi}2\psi\gamma^0\gamma^1\partial_1\psi-\frac{m}2\psi\gamma^0\psi
\right\}.\label{ham}\ee
This resembles $\phi^4$ theory, and for the same reason, 
we cannot calculate the $\mathcal{C}$ operator perturbatively.  

\subsection{Functional integral}
It may be useful to rederive the Hamiltonian using the functional
integral approach.  The partition function is
\be
Z=\int[\rmd \psi][\rmd \phi]\,\rme^{\rmi\int\rmd t\,\rmd x\, L},
\ee
where in the Coulomb gauge (with $A^0=\phi$)
\be
L=\frac12(\partial_1\phi)^2-\rmi j^0\phi-\frac12\psi\gamma^0\gamma^\mu
\frac1\rmi\partial_\mu\psi-\frac{m}2\psi\gamma^0\psi.
\ee
We integrate out the scalar potential $\phi$ by completing the square,
\be
\frac12(\partial_1\phi)^2-\rmi j^0\phi=\frac12\left(\partial_1\phi+
\frac\rmi{\partial_1}j^0\right)^2-\frac12j^0\frac1{\partial_1^2}j^0,
\ee
where 
\be
\frac1{\partial_1^2}=\frac12|x-y|.
\ee
The functional integral on $\phi$ is carried out over real values
of $\phi(x)$.
Then, the partition function can be written as
\be
Z=\int[\rmd \psi]\rme^{\rmi\int \rmd t\,\rmd x(\pi\dot\psi-\mathcal{H}
)},\label{pf}
\ee
where the momentum conjugate to $\psi$ is
\be
\pi=\frac{\partial L}{\partial\dot\psi}=\frac{i}2\psi.
\ee
The result for $H$, given in \eref{ham}, is reproduced.
Because the sign of the quartic term in $H$ is reversed, presumably
we can no longer regard $[\rmd \psi]$ in \eref{pf} as over ``real''
values of $\psi$.  Since Grassmann integration is a formal procedure,
it is not immediately clear how to proceed.
Henceforth, we will set the fermion mass $m=0$, so we will refer to the
Schwinger model proper.
\subsection{ETCR of currents}
It is easy to check that
\be
[j^0(x,t),j^0(y,t)]=0.\label{j0j0}\ee
However, it requires a point-splitting calculation 
[which does not modify \eref{j0j0}] to verify that
\be
[j^0(x,t),j^1(y,t)]=-\frac{\rmi e^2}\pi\frac\partial{\partial x}\delta(x-y).
\label{j0j1comm}\ee
The key element in the latter is that the singular part of the 2-point fermion
correlation function is given by the free Green's function:
\numparts
\bea
\langle \psi_\alpha(x)(\psi(y)\gamma^0)_\beta\rangle
=\frac1\rmi G_{\alpha\beta}(x-y),\\
\quad G(z)=-\frac1{2\pi}\frac{\gamma_\mu z^\mu}{z^2+\rmi\epsilon}.\eea
\endnumparts
This agrees with the massless  fermion propagator found in
\eref{gee} below.
\subsection{Conservation of electric charge}
The electric current is exactly conserved:
\numparts
\be
\partial_0 j^0=\frac1\rmi[j^0,H]=-\partial_1 j^1,\ee
or
\be
\partial_\mu j^\mu=0.\ee
\endnumparts
\subsection{Axial-vector anomaly}
In 2-dimensions, the dual current is
\be
{}^*j^\mu=\epsilon^{\mu\nu}j_\nu,\quad {}^*j^0=j_1,\quad {}^*j^1=j^0.\ee
Now, using the above commutator between $j^0$ and $j^1$, we find
\bea
\partial_0{}^*j^0&=&\partial_0j_1=\frac1\rmi[j_1,H]\nn\\
&=&-\partial_1j^0+\frac1\rmi\left[j_1(x),\frac14\int \rmd y\,
\rmd z\, j^0(y)|y-z|j^0(z)
\right],\eea
so from \eref{j0j1comm} this can be rewritten as
\bea
\partial_\mu{}^*j^\mu(x)&=&-\frac{e^2}{2\pi}\int \rmd y\,\rmd z
\,\partial_x\delta(x-y)|y-z|j^0(z)\nn\\
&=&-\frac{\rmi e^2}\pi\partial_x A^0=\frac{\rmi e^2}\pi E.\eea
This is the two-dimensional version of the famous
Schwinger-Adler-Bell-Jackiw anomaly \cite{anomaly}.

\subsection{Schwinger mass generation}
Combine the current conservation and axial-current non-conservation:
\numparts
\bea
\partial_1[\partial_0j^0+\partial_1 j^1&=&0]\\
\partial_0\big[\partial_0j^1+\partial_1 j^0&=&\frac{\rmi e^2}\pi E\big],\eea
\endnumparts
together with the Maxwell equation \eref{max}
to obtain ($\partial^2=-\partial_0^2+\partial_1^2$)
\be
\left(\partial^2+\frac{e^2}\pi\right)j^1=0.\ee
This corresponds to a {\em spacelike singularity}, a pole at
\be
k^2=-\partial^2=\frac{e^2}\pi,\ee
implying complex energies!

\subsection{Perturbation theory}
This result is consistent with perturbation theory, where in general
we expect all we have to do is replace
\be
e\to \rmi e.\ee
In fact, the Schwinger mass comes from one-loop vacuum polarization.
In particular, the $\mathcal{C}$ operator appears to have no effect
on the weak-coupling expansion \cite{Bender:2005sc}.  In general, it appears
only ephemerally in the functional integral formulation \cite{Jones:2006sj}.

\begin{figure}
\centering
\epsfig{file=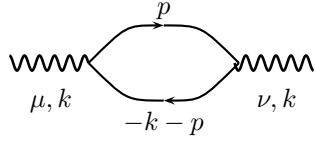}
\caption{Lowest-order vacuum polarization graph}
\label{fig4.loop}
\end{figure}

Let us sketch the Feynman diagrammatic argument for the mass generation
mechanism in the modified Schwinger model.  That is, we calculate the
vacuum polarization operator for massless QED in lowest order as shown
in \fref{fig4.loop}.  This corresponds to
\be
\Pi^{\mu\nu}=-(\rmi e)^2\,\mbox{tr}\int\frac{(\rmd^d p)}{(2\pi)^d}
\gamma^\mu\frac{-\gamma p}{p^2}
\gamma^\nu\frac{\gamma(k-p)}{m^2+(p-k)^2},
\ee where we have regulated the integral by working in $d$ dimensions.
The trace is evaluated as
\be
\mbox{tr}\gamma^\alpha\gamma^\beta\gamma^\gamma\gamma^\delta
=d\left(g^{\alpha\beta}g^{\gamma\delta}+g^{\alpha\delta}g^{\beta\gamma}-
g^{\alpha\gamma}g^{\beta\delta}\right).
\ee
The denominators are combined according to
\be
\frac1{p^2(p-k)^2}=\int_0^\infty \rmd s\,s\int_0^1 \rmd  u\,\rme^{-\rmi s\chi},
\ee
with
\be
\chi=(1-u)p^2+u(p-k)^2=(p-ku)^2+k^2u(1-u).
\ee
Now the integration variable is shifted, $p\to p+ku$, and odd terms disappear
upon symmetric integration, leaving us with
\bea
\Pi^{\mu\nu}=
-(\rmi e)^2d\int_0^\infty 
\rmd s\,s\int_0^\infty \rmd u\int\frac{\rmd^d p}{(2\pi)^d}
\bigg[2p^\mu p^\nu-2k^\mu k^\nu u(1-u)\nn\\
\qquad\mbox{}-g^{\mu\nu}\left(p^2-k^2u(1-u)\right)
\bigg]\rme^{-sp^2}\rme^{-su(1-u)k^2}.
\eea
Note that in $d=2$ dimensions, the contraction (trace) of the tensor, $\Pi^\mu
{}_\mu$, vanishes, which indicates that, apparently, the only gauge-invariant
result could be zero.  That this is incorrect is the result of the
quantum anomaly, which appears only by setting $d=2$ at the end of the
calculation. Proceeding onward, we use the following 
momentum integrals:
\numparts
\bea
\int \rmd^d p \,\rme^{-sp^2}=\left(\int\rmd p\,\rme^{-sp^2}\right)^d=\left(
\frac{\pi}{s}\right)^{d/2},\\
\int \rmd^d p\,p^\mu p^\nu \rme^{-sp^2}=-g^{\mu\nu}\frac1d \frac\rmd{\rmd s}
\int \rmd^dp\,\rme^{-sp^2}=\frac1{2s}\left(\frac{\pi}{s}\right)^{d/2}
g^{\mu\nu}.
\eea
\endnumparts
When the $s$ integrals are now carried out, we obtain an explicitly 
gauge-invariant form:
\be
\fl\Pi^{\mu\nu}=-2^{1-d}(\rmi e)^2\pi^{-d/2}d\Gamma(2-d/2)\int_0^1 \rmd u
\left[k^2u(1-u)\right]^{d/2-1}
\left(g^{\mu\nu}-\frac{k^\mu k^\nu}{k^2}\right),
\ee
which is divergent for $d=4$, but finite for $d=2$:
\be
\Pi^{\mu\nu}=-\frac{(\rmi e)^2}\pi\left(g^{\mu\nu}-\frac{k^\mu k^\nu}{k^2}\right).\ee

We can similarly calculate the correction to the fermion propagator,
given by the mass operator, using the covariant photon propagator,
\be
\Sigma=(\rmi e)^2\int\frac{\rmd^d k}{(2\pi)^d}\frac{\gamma^\mu \gamma(p-k)
\gamma^\nu}{k^2(p-k)^2}\left(g_{\mu\nu}-\frac{k_\mu k_\nu}{k^2}\right).
\ee
The gamma matrix structures are reduced as
\numparts
\bea
\gamma^\lambda\gamma(p-k)\gamma_\lambda=(d-2)\gamma(p-k),\\
\gamma k\,\gamma(p-k)\,\gamma k=k^2\gamma(p-k)+\gamma k [(p-k)^2+k^2-p^2],
\eea
\endnumparts
and so the integral, after the substitution $k\to k+pu$, can be evaluated
as
\bea
\fl\gamma p\int_0^\infty \rmd s\,s\int_0^1 \rmd u\left(\frac\pi{s}\right)^{d/2}
\left[(d-3)(1-u)-u-p^2\frac\rmd{\rmd p^2}\right]\rme^{-su(1-u)p^2}\nn\\
\fl=\gamma p\int_0^1 \rmd u\left(\frac{d}2-1\right)(1-2u)\pi^{d/2}\Gamma(2-d/2)
[p^2u(1-u)]^{d/2-2}\nn\\
\fl=\gamma p\,
\pi^{d/2}\Gamma(2-d/2)(p^2)^{d/2-2}\left(\frac{\Gamma(d/2)\sqrt{2\pi}}
{\Gamma(d/2-1/2)2^{d-5/2}}-2\frac{\Gamma(d/2)^2}{\Gamma(d-1)}\right)\to0,
\eea
as $d\to2$, that is, the mass operator vanishes.

In the conventional theory, the vacuum polarization, iterated, yields the
boson mass generated in the Schwinger model:
\be
D(k)=\frac1{k^2}-\frac1{k^2}\frac{e^2}\pi\frac1{k^2}+\dots
=\frac1{k^2+e^2/\pi},
\ee
while there is no correction to the fermion propagator,
\be
\rmi
G(p)=\frac1{\gamma p}.
\ee
The corresponding iterated one-loop propagators in coordinate space are
\numparts
\bea
D(x)=\frac1{2\pi}K_0(mx), \\
G(x)=-\frac1{2\pi}\frac{\gamma^\mu x_\mu}{x^2},\label{gee}
\eea
\endnumparts
where in the conventional theory $m^2=e^2/\pi$, while
the mass-squared 
is reversed in sign in the $\mathcal{PT}$ theory.  These propagators
are given
in terms of the Euclidean distance $x=\sqrt{x^2}$.
Of course, in other gauges, there are corrections to the fermion
propagator.
However, in Schwinger's words \cite{Schwinger:1962tp}, 
``the detailed physical interpretation
of the Green's function is rather special and apart from our main purpose.''
\section{Conclusions}

Perturbation theory evidently fails to give a positive spectrum
to the massless $\mathcal{PT}$-symmetric electrodynamics in 2 dimensions.
More generally, this reflects the lack of unitarity of the $S$-matrix
for the theory.  This phenomenon has already been noticed by other
authors: For example, Jones \cite{jones} shows for localized non-Hermitian
potentials probability is not conserved, unless the Hilbert space
metric is changed; and Smilga \cite{Smilga:2007mq}
shows that the $S$-matrix of a non-Hermitian
theory defined in terms of ``transition amplitudes between {\em conventional\/}
asymptotic states is not unitary.''

Therefore, it seems that nonperturbative effects (strong field effects) 
presumably resolve this issue. It is necessary to do more than merely
compute the $\mathcal{Q}$ operator in field theory, which as we have seen
makes no explicit appearance in the functional or perturbative formulation
of the theory, but we must determine the asymptotic states, presumably
defined dynamically.  This is a formidable problem.

Previous work on these questions has concentrated on quantum-mechanical
and scalar-field theory examples.
Clearly there are issues unsolved relating to fermions and gauge theories
in the $\mathcal{PT}$-context.  In particular, the formal technique of
Grassmann integration needs to be redeveloped.
One must find ways to reformulate
usual field theoretic tricks, such as bosonization.

Another illustration of the failure of perturbation theory to capture
the essence of a $\mathcal{PT}$ theory is discussed in the Appendix.

\appendix
\section{Zero-Dimensions}
We contrast the zero-dimensional partition functions for a
conventional and a $\mathcal{PT}$-symmetric $x^{2+N}$ theory:
\numparts
\bea
Z_N^c(K)&=&\int_{-\infty}^{\infty} \rmd x\,\rme^{-x^2- g x^{2+N}-Kx},\\
Z_N(K)&=&\int_C \rmd x\,\rme^{-x^2-gx^2(\rmi x)^N-Kx}.\eea
\endnumparts
Such examples were discussed earlier in \cite{Bender:1999ek}.
The contour $C$ in the latter integral
 is taken in the lower half plane, along Stokes wedges centered
on the lines
\be
\theta=-\frac{N\pi}{4+2N},\quad \theta=\pi-\frac{N\pi}{4+2N},
\ee
which have width $2\pi/(4+2N)$, so that the
integrand decays exponentially fast.

Note that the $\mathcal{PT}$-symmetric theory has a perturbation series
that doesn't appear to know about the path of integration:
\bea
Z_N(K)&=&\sqrt{\pi}\exp\left[g\left(-\rmi\frac{\rmd}{\rmd K}
\right)^{N+2}\right]\rme^{K^2/4}
\nn\\
&=&\sqrt{\pi}\rme^{K^2/8}\sum_{n=0}^\infty \left(\frac{(-1)^N g}{2^{1+N/2}}
\right)^n\frac1{n!}D_{n(N+2)}\left(\frac{\rmi K}{\sqrt{2}}\right),
\label{zexp}\eea
where $D_m(x)$ is the parabolic cylinder function.

When we set $N=2$ and $K=0$ we get the
$-x^4$ theory without sources, 
for which we have the closed form for the vacuum amplitude
\be
Z_2(0)=\frac{\pi}{4\sqrt{g}}\rme^{-1/8g}\left[I_{1/4}\left(
\frac1{8g}\right)
+I_{-1/4}\left(\frac1{8g}\right)\right].\label{zpt}\ee
The conventional theory in the same situation has the closed form
\be
Z_2^c(0)=\frac1{2\sqrt{g}}\rme^{1/8g}K_{1/4}\left(\frac1{8g}\right).\label{zconv}
\ee

Directly from \eref{zpt}, or from the previous expansion
\eref{zexp}, we find the weak-coupling
expansion ($g\to0$)
\be
Z_2(0)\sim\sqrt{\pi}\left(1+\frac34 g+\frac{105}{32} g^2+\dots\right);
\ee
the expansion of $Z_2^c$ differs only in the sign of $g$.
The conventional theory is Borel summable, while the $\mathcal{PT}$ series
is not.

However, the correspondence is not so simple in strong-coupling.
Even the leading prefactors in the strong-coupling expansions 
 are different: ($g\to\infty$)
\numparts
\bea
Z_2^c(0)&\sim& \frac{\sqrt{2}\pi}{2g^{1/4}\Gamma(3/4)}\left[1-\frac1{4\sqrt{g}}
\frac{\Gamma(3/4)}{\Gamma(5/4)}+\dots\right],\\
Z_2(0)&\sim& \frac{\pi}{2g^{1/4}\Gamma(3/4)}\left[1+\frac1{4\sqrt{g}}
\frac{\Gamma(3/4)}{\Gamma(5/4)}+\dots\right].\eea
\endnumparts

\ack
This work is supported in part by grants for the US Department of Energy
and from the US National Science Foundation.  We thank all the participants
in the 6th International 
Workshop on Pseudo-Hermitian Hamiltonians in Quantum Physics for
their comments and cross-fertilization, and are particularly indebted to
Carl Bender for his assistance.

\end{document}